\documentclass[twocolumn,aps,showpacs,raggedbottom,nobalancelastpage,
amsmath,amssymb,superscriptaddress]{revtex4}
\usepackage{graphicx}
\usepackage{natbib}
\usepackage{color} 
\newcommand{\beq}{\begin{equation}}
\newcommand{\eneq}{\end{equation}}
\newcommand{\be}{\begin{equation}}
\newcommand{\ee}{\end{equation}}
\newcommand{\bea}{\begin{eqnarray}}
\newcommand{\eea}{\end{eqnarray}}

\begin{document}

\title{Anomalous Josephson Effect in S/SO/F/S heterostructures}
\author{M.Minutillo}
\affiliation{Dipartimento di Fisica "Ettore Pancini", Universit\`a di Napoli ``Federico II'', 
Monte S.Angelo, I-80126 Napoli, Italy}

\author{D. Giuliano}
\affiliation{Dipartimento di Fisica, Universit\`a della Calabria, Arcavacata di Rende, I-87036, Cosenza, Italy}  
\affiliation{INFN, Gruppo 
Collegato di Cosenza, Arcavacata di Rende, I-87036, Cosenza, Italy}

\author{P. Lucignano}
\affiliation{CNR-SPIN, Monte S.Angelo -- via Cinthia,  I-80126 Napoli, Italy}
\affiliation{Dipartimento di Fisica "Ettore Pancini", Universit\`a di Napoli ``Federico II'', 
Monte S.Angelo, I-80126 Napoli, Italy}

\author{A. Tagliacozzo}
\affiliation{Dipartimento di Fisica "Ettore Pancini", Universit\`a di Napoli ``Federico II'', 
Monte S.Angelo, I-80126 Napoli, Italy}
\affiliation{CNR-SPIN, Monte S.Angelo -- via Cinthia,  I-80126 Napoli, Italy}
\affiliation{INFN, Gruppo 
Collegato di Cosenza, Arcavacata di Rende, I-87036, Cosenza, Italy} 

\author{G. Campagnano}
\affiliation{Dipartimento di Fisica "Ettore Pancini", Universit\`a di Napoli ``Federico II'', 
Monte S.Angelo, I-80126 Napoli, Italy}
\affiliation{CNR-SPIN, Monte S.Angelo -- via Cinthia,  I-80126 Napoli, Italy}
\begin{abstract}
 We study the anomalous Josephson effect, as well as the dependence on the direction 
 of the critical Josephson current, in an S/N/S junction, where the normal part is realized by alternating 
 spin-orbit coupled and ferromagnetic layers. We show that to observe these effects it is sufficient 
 to break  spin rotation and time reversal symmetry in spatially separated regions of the junction.
 Moreover, we discuss how to further improve these effects by engineering multilayers structures 
 with more that one couple of alternating layers.     

\end{abstract}

\maketitle

\section{Introduction}

 A continuosly growing interest has recently arisen in mesoscopic systems in which  conventional superconductivity, spin orbit interaction, 
and magnetism come into play at the same time. For instance, Josephson junctions realized with semiconducting nanowires made with group III-V semiconductors, 
such as  InAs or InSb (which are chosen because of their strong spin orbit coupling and large g factor\cite{nanowire_review_1,nanowire_review_2}) 
have attracted  much attention as possible platform to support topologically protected Majorana states \cite{alicea_1}. Also,  higher-periodicity 
junctions have been proposed as arising from the combined effects of topology and electronic correlations  \cite{fegel,gs1}
Parallel to the search for  topologically protected states these systems have also shown to be an ideal playground to investigate non conventional Josephson effects, 
such as the Anomalous Josephson Effect\cite{Yokoyama:2014} (AJE), which is the main topic of this paper. 

In its standard form the  dc Josephson current flowing between two superconducting electrodes at a fixed phase difference $\varphi$
is expressed via  a sinusoidal current-phase relation (CPR) given by \cite{Barone} $I(\varphi)=I_c \sin \varphi$, with the 
critical current $I_c$ representing the maximum non dissipative 
current that the Josephson junction can support. Among the specific features of the CPR above, one has to stress  that:\\
{\em i)} the current is strictly zero for $\varphi=0,\pi$, \\
{\em ii)} the critical current {\em does not} depend on the current direction.\\
In general, it has been shown that, when a system exhibits either  time reversal symmetry, or spin rotational symmetry (or both), 
$I ( \varphi )$  must necessarily be equal to zero for $\varphi=0, \pi$. Therefore, in order to find AJE as an anomaly in the CPR, that is, 
to have $I ( \varphi_0 ) = 0$ at $\varphi_0 \neq 0 , \pi$,   
one needs to break simultaneously these two symmetries \cite{Yokoyama:2013,Campagnano:2015}. 

The AJE has been initially predicted in systems with non-conventional superconductivity\cite{Larkin:1986, Satoshi:2000, 
Sigrist:1998,PhysRevB.52.3087, PhysRevLett.102.227005, PhysRevLett.99.037005}. Further studies have shown that there is a
large group of systems which might exhibit the AJE, in particular S/N/S junctions where the normal region is: a magnetic normal 
metal\cite{Buzdin:2008,PhysRevB.67.184505,Nat_phys_Eschrig_2008,PhysRevB.76.224525,PhysRevLett.98.077003,PhysRevLett.102.017001}, 
a one-dimensional quantum wire, a quantum dot\cite{Zazunov:2009,brunetti:2013}, a multichannel system with a barrier or a quantum 
point contact\cite{Reynoso:2008,Reynoso:2012}, a semiconducting nanowire\cite{Yokoyama:2013,Yokoyama:2014}. Anomalies of the 
Josephson current have also been predicted  in presence of Coulomb interactions  and spin orbit interaction (SOI)  for a wire \cite{Krive:2004,Krive:2005}  
or a Quantum Dot \cite{brunetti:2013} contacted with conventional superconductors. Closely related to our a work is a recent proposal 
suggesting the possibility of obtaining a $\varphi_0$-junction by means of a non-coplanar ferromagnetic junction \cite{Silaev:2017}. 
Remarkably, the AJE can also be exploited to discern topological versus conventional 
superconductivity \cite{Nesterov:2016,Marra:2016,Schrade:2017}. 

On the experimental side, a nonzero shift $\varphi_0$ has been recently  demonstrated 
using a gated InSb nanowire embedded in a superconducting quantum interference device\cite{Szombati:2016}.  Even more interestingly, 
some systems exhibit the remarkable feature that the anomalous CPR ($\varphi_0 \neq 0 , \pi $) is 
 accompanied by a direction dependent critical current, that is, by an asymmetry $I_{c+}-I_{c-} \neq 0$, with 
 $I_{c+}$ and $I_{c-}$ respectively corresponding to the absolute value of the maximum and of the minimum value reached by  
 $I ( \varphi )$.

In this article we study the possibility to observe the anomalous Josephson effect and the direction dependent critical current in an S/N/S junction, 
with the N part realized with an heterostructure composed by two or four layers where  a spin orbit coupled  region is alternated to a ferromagnetic one. 
Our proposal is motivated by the observation that separating in space the spin orbit coupled region(s) from the ferromagnetic one(s) is expected to offer 
some advantages with respect to the ''standard'' approach, in which one applies an external magnetic field to a material with 
large spin-orbit. Indeed, in our  case  one might use a material with large spin orbit coupling which might not have a large $g$-factor and hence 
requiring magnetic field too large to be sustained by the superconducting leads.

The article is organized as follows: in section \ref{model} we introduce our model and discuss how to compute the Josephson current from the scattering 
matrix of the normal region.
In section \ref{results} we present and discuss our result for the two-layers (section \ref{results_A}) and the four-layers (section \ref{results_B}) normal region. 
In section \ref{ramaan} we  
present a random matrix analysis to justify why we need to consider at least two transport 
channels to look for a large  asymmetry. 
We summarize our findings  and provide our conclusions in section \ref{summary}. 
In appendix \ref{scatmay} we provide details about how to compute the scattering matrices of the different layers constituting the 
normal region and how to combine them to construct the scattering matrix of the whole normal region.

\section{Model and calculation of the Andreev bound states}
\label{model}

\begin{figure}[ht]
\includegraphics[width=.8\linewidth]{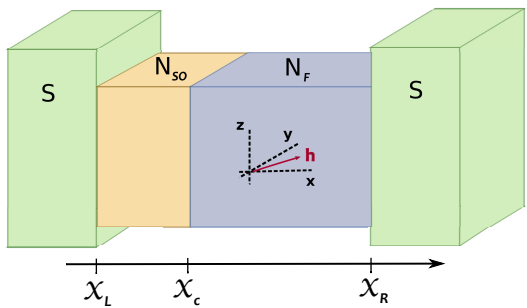}
\caption{Schematic representation of the device. The normal region consists of a heterostructure made by region with spin-orbit coupling ($N_{SO}$) connected to a ferromagnetic region ($N_F$). 
The magnetization (or alternatively a magnetic field) is assumed to be in the $xy$ plane.
For the sake of simplicity in the calculations we consider a two-dimensional system.}
\label{scheme_2_reg}
\end{figure}

In Fig.~\ref{scheme_2_reg} we present a scheme of our setup. As discussed in the following, 
we model our system as   a quasi one-dimensional heterostructure connected to two  conventional s wave 
superconductors to form a S/$N_{SO}$/$N_{F}$/S junction.  We assume that a strong Rashba SOI is present in the $N_{SO}$ region,  
while the region $N_{F}$ is characterized by an exchange field $\vec{h}$  or alternatively by an externally applied magnetic field.
In order to avoid unnecessary complications, we assume that the effective electronic mass is the same in all the different regions.
Nevertheless,  our analysis can be easily generalized to the case of different effective masses. In addition, 
we assume that the SOI is zero  in the superconducting leads, since we want to focus onto the case of non-topological 
superconducting leads. In fact, junctions between topological superconductors and normal 
wires are relevant for the physics of emerging real Majorana fermionic modes \cite{fidk,giuaf_1,giuaf_2}, but not 
for AJE, which is what we focus on in our work.

It has been previously pointed out that   the AJE is maximum when the 
magnetic field (or the magnetization) is parallel to the effective  spin-orbit (SO) field, which corresponds to the ''effective magnetic field'' due to the SOI 
\cite{Szombati:2016,Rasmussen:2016}.
For this reason, since we are interested in configurations that maximize $\varphi_0$, in the 
following we  consider only the case of an in-plane magnetic field (or magnetization). This 
implies that in this case there are no magnetic orbital effects and, accordingly, 
only the Zeeman coupling has to be properly  taken into account.

In order to compute the Josephson current, we look for  solutions of the Bogoliubov-de Gennes equations:
\begin{multline}
\label{bdg}
\mathcal{H}_{BdG} \left(
\begin{array}{c}
{\bf u}(x,y) \\ {\bf v}(x,y)
\end{array}
\right)
= \epsilon 
\left(
\begin{array}{c}
{\bf u}(x,y) \\ {\bf v}(x,y)
\end{array}
\right) \; , 
\end{multline}
with 

\begin{equation}
 \mathcal{H}_{BdG}  = \left(
\begin{array}{cc}
H-E_F & \Delta \\
 \Delta^\dag & -(H^*-E_F)
\end{array}
\right)
\; . 
\label{bdg.2}
\end{equation}
\noindent
In Eqs.(\ref{bdg}),  $\epsilon$ measures the energy with respect to the Fermi level $E_F$, while ${\bf u}(x,y)$ and ${\bf v}(x,y)$ are
respectively the electron and hole spinors in the Nambu representation. To model the junction, we take 
the s-wave pairing potential to be given by  
\begin{equation}
\hat{\Delta}=\Delta(x)\left(
\begin{array}{cc}
0 & -1 \\
1 & 0
\end{array}
\right)\:,
\end{equation}
with 
\begin{equation}
\Delta(x)=\Delta_0 \left[  \Theta(x_L-x)e^{-i \varphi/2}+\Theta(x-x_R)e^{i \varphi/2} \right]
\: . 
\label{pairing}
\end{equation}
\noindent
In the normal Hamiltonian $H$ in Eqs.\ref{bdg.2}, we assume that the electrons are free to propagate in the $x$ 
direction, while we introduce an harmonic confining potential in the $y$ directions,  which comes out to be a 
 particularly  convenient choice, when expressing the matrix elements of the SOI operator 
 \cite{Governale:2002,marigliano:2004}. The corresponding Hamiltonian reads:

\begin{multline}\label{hamiltonian}
H=\frac{p_x^2}{2m}+\frac{p_y^2}{2m}+\frac{1}{2}m \omega^2 y^2 +\frac{\alpha(x)}{\hbar}(\sigma_x p_y-\sigma_y p_x)
\\ +h(x) \hat{n}\cdot \vec{\sigma}+\frac{i}{2}\partial_x \alpha(x) \sigma_y  \, ,
\end{multline}
where 
\begin{equation}
\alpha(x)=
\left\{\begin{array}{cc}
\alpha_0 & \mbox{for} \, \, x_L<x<x_c \\
0 & \mbox{otherwise}
\end{array}
\right.
\end{equation}
with $\alpha_0$ being the strength of the Rashba SOI, and $\vec{h} = \hat{n} h ( x )$, with $ | \hat{n} | = 1$ and 
\begin{equation}
h(x)=
\left\{\begin{array}{cc}
h_0 & \mbox{for} \, \,  x_c<x<x_R \\
0 & \mbox{otherwise}
\end{array}
\right.
\end{equation}
with $h_0$ being the intensity of the exchange field. In Eq.~\ref{pairing} $\Theta (x)$ is the Heaviside step function,  
which corresponds to a rigid, non self-consistent, profile for 
the pairing term (see e.g. Ref.~\onlinecite{Likharev} for a discussion of the physical applicability
of the model with stepwise changes in the physical parameters as a
function of the position). In addition, without loss of 
generality, we set the phase difference $\varphi$ to be symmetrically 
distributed between the two superconducting leads. 

The spectrum of Eq.~\ref{bdg} consists of a finite set of bound states (Andreev levels) with energy $|\epsilon|<\Delta_0$,
and a continuum of states with $|\epsilon|>\Delta_0$. The current can be obtained from the free 
energy $F(\varphi)$ by the thermodynamic relation\cite{Bloch:1970} 
\begin{equation}\label{curr-free-energy} 
I(\varphi)=\frac{2e}{\hbar}\frac{d F}{d\varphi} \; , 
\end{equation}
\noindent
with the free energy in Eq.~\ref{curr-free-energy} obtained by considering contributions from all 
the states in the spectrum. 

In this work we only consider the short junction limit, in which case only the subgap Andreev states contribute to the Josephson current
(the complementary long junction limit can be addressed by means, for instance, of the techniques developed in 
Refs.~\onlinecite{giuaf_j1,giuaf_j2,nava_ring}). 
Moreover, we limit our analysis to the zero temperature case, which allows for simplifying Eq.~\ref{curr-free-energy} to:

\begin{equation}
I(\varphi)=\frac{e}{\hbar}\sideset{}{'}\sum_n  \frac{\partial E_n(\varphi)}{\partial \varphi}.
\label{cur1}
\end{equation}
In Eq.~\ref{cur1}  $n$ labels the Andreev states, whose 
energies correspond to  the discrete spectrum of Eq.~\ref{bdg}, and the primed sum means that only negative energy (occupied) Andreev states 
are  considered (Notice that, in Eq.~\ref{cur1}, there is a factor 2 missing, with respect 
Eq.~\ref{curr-free-energy}. In fact, this takes into account that, due to the lack of spin conservation, 
because of SOI, the  spin  degeneracy in the counting of Andreev levels is lifted in Eq.~\ref{cur1}).
As we model the nanowire by means of a  transverse harmonic confining potential in  the y-direction, while the electrons 
propagate as free particles in the x-direction, we may  derive the Andreev states by employing the  scattering matrix approach put
forward in  Ref.~\onlinecite{Beenakker:1991}. Specifically,  
one can ideally think of the incoming and outgoing scattering 
states on the normal region as respectively the outgoing and incoming states at the 
superconducting regions. The Andreev bound states correspond to the stationary solutions bound 
within the normal region and, accordingly, they are 
described as evanescent waves in the superconducting leads.
At energies below the superconducting gap $\Delta_0$,
at the interface between the normal and the superconducting regions only
intra-channel Andreev scattering takes place 
where a hole  (electron) with spin $\sigma$
is reflected as an electron (hole) with spin $-\sigma$.
These processes are encoded in the 
relations

\begin{equation}
\left(\begin{array}{c}
a_{eL} \\
a_{eR} \\
a_{hL} \\
a_{hR}
\end{array}
\right) =\hat{S}_A
\left(\begin{array}{c}
b_{eL} \\
b_{eR} \\
b_{hL} \\
b_{hR}
\end{array}
\right) \; .
\label{asm}
\end{equation}
\noindent
The Andreev scattering matrix $\hat{S}_A$ is defined as:
\begin{equation}
\hat{S}_A = 
\left(\begin{array}{cc}
0 &  \hat r_{eh}\\
\hat r_{he}    & 0  
\end{array}
\right) \; , 
\end{equation}
with 
\begin{equation}
\hat r_{eh} =  i\,
e^{-i \gamma }\left(\begin{array}{cc}
 \hat 1 \otimes \hat \sigma_y e^{- i \varphi/2}&  0\\
0    &  \hat 1 \otimes \hat \sigma_y e^{+ i\varphi/2}
\end{array}
\right)\, ,
\end{equation}
and
\begin{equation}
\hat r_{he} = - i\,
e^{-i \gamma }\left(\begin{array}{cc}
 \hat 1 \otimes \hat \sigma_y e^{+i \varphi/2}&  0\\
0    &  \hat 1 \otimes \hat \sigma_y e^{- i\varphi/2}
\end{array}
\right)\, .
\label{ll.1}
\end{equation}
In Eq.~\ref{ll.1}  $\hat 1$ is the identity matrix in the channel space, 
the $\hat \sigma_y$ Pauli matrix acts in 
the spin space, and $\gamma=\arccos(\epsilon/\Delta_0)$. 
In the normal region there is no conversion of electron into hole states
but only normal scattering processes are  allowed. The corresponding 
scattering matrix is purely normal,  implying that there are no off-diagonal 
terms corresponding to scattering of particles into holes, and vice versa.
Whithin the central region, this  allows us to write
\begin{equation}
\left(\begin{array}{c}
b_{eL} \\
b_{eR} \\
b_{hL} \\
b_{hR}
\end{array}
\right)
=\left(
\begin{array}{cc}
\hat{S}_e(\epsilon) & {\bf 0} \\
{\bf 0} & \hat{S}_h(\epsilon)  
\end{array}
 \right) 
 \left(\begin{array}{c}
a_{eL} \\
a_{eR} \\
a_{hL} \\
a_{hR}
\end{array}
\right),
\label{normal}
 \end{equation}
 \noindent
 with $\hat{S}_e(\epsilon) $ and $ \hat{S}_h(\epsilon) $ being the normal scattering 
 matrix for particles into particles and for holes into holes, respectively. 
The energy of the Andreev bound states is determined by the 
secular equation \cite{Beenakker_universal}
\begin{equation}\label{determinant}
\det \left[ \hat{1} -\hat{r}_{eh} \hat{S}_{h}(-\epsilon)  \hat{r}_{he}   \hat{S}_{e}(\epsilon) \right]=0.
\end{equation}
In the short-junction limit case the Thouless energy $E_c\simeq\hbar/\tau_{dwell}$ 
(with $\tau_{dwell}$ the dwell time in the junction)  is much larger than the superconducting gap $\Delta_0$, 
in this case one can safely disregard the energy dependence of the scattering matrix 
and take $ \hat{S}^*_{h}(-\epsilon)=\hat{S}_{e}(\epsilon)\simeq\hat{S}_{e}(0) $. 
Therefore, in order to solve Eq.~\ref{determinant}, one only needs to calculate 
the scattering matrix of the normal region at the Fermi energy.
The approximation above allows for a further simplification in the calculation of the Andreev spectrum.
Indeed we can introduce the matrix $\hat{W}=\exp(2i\gamma)\hat{r}_{eh} \hat{S}^*_{e}(0)  \hat{r}_{he} \hat{S}_{e}(0)$,
which is unitary, with a set of eigenvalues $\{w_i\}$ of  modulus one.
 Using Eq.~\ref{determinant} one sees that, in terms of the $\{w_i\}$,
 the Andreev levels are then obtained from the relation 
\begin{equation}
\arccos(\frac{\epsilon}{\Delta_0})=\frac{1}{2}\arg(w_i) \,.
\label{keyeq}
\end{equation}
\noindent
Eq.~\ref{keyeq} is what we  have been using in the following to compute the Andreev energy levels and to 
accordingly compute the Josephson current. As, within our assumptions on the  model Hamiltonian we 
use, the key ingredient determining Eq.~\ref{keyeq} is the normal region scattering 
matrices $\hat{S}_{e , h} ( \epsilon )$, we outline 
the details of their derivation in appendix \ref{scatmay}. 

\section{Results and discussion}
\label{results}

We now present our main results by displaying $ I ( \varphi )$ as a 
function of $\varphi$ calculated using Eqs.~\ref{cur1} and ~\ref{determinant}, 
for  several representative values of the SOI and of  $h_0$.

\subsection{2 Regions}
\label{results_A}

Here we consider the case described by the Hamiltonian of Eq.~\ref{hamiltonian}, where the normal region consists of two 
regions; the first one characterized by a SOI $\alpha$ and a second one characterized by an exchange field $\vec{h}$ 
by also assuming, for the sake of simplicity,   a perfect transparency between the 
two regions.   We consider first the case of spin-orbit region coupled to a ferromagnetic region and 
study the CPR for a fixed value of $\alpha$ and  several values of the exchange field. As it was shown in Ref.~\onlinecite{Szombati:2016}, when the magnetic field  is perpendicular to the SO field  (
which corresponds to the ''effective magnetic field'' due to the SOI and, given the confinement in the y direction and the assumptions above, in our case 
is directed along the y axis), no anomaly
in the CPR is observed. This is perfectly consistent with the plots we show in Fig.~\ref{current-phase-2},
where we assume $\theta=0$, with $\theta$ being the angle between the (in-plane) magnetization 
$\vec{h}$ and the x axis, that is, $\vec{h} / | \vec{h} | \equiv \hat{n} = ( \cos ( \theta ) , \sin ( \theta ) , 0)$ -- see appendix \ref{scatmay}  for
details.  To spell our why the anomaly is zero when $\theta=0$, let 
us consider the unitary operator $\mathcal{O}$, defined as 
$ \mathcal{O} = i \mathcal{K} {\bf \Pi}_y$,
with ${\bf \Pi}_y$ being the parity operator in the y direction, that is, 
${\bf \Pi}_y  y {\bf \Pi}_y^{-1} = - y $. By direct calculation, one readily checks that,
with the system Hamiltonian $\mathcal{H}_{BdG}(\varphi)$ in Eq.~\ref{bdg} (and, more 
generically, with any Hamiltonian envisaging a parabolic confinement in the y direction), 
one obtains $\mathcal{O} \mathcal{H}_{BdG}(\varphi ; \theta )\mathcal{O}^{-1}=\mathcal{H}_{BdG}(-\varphi , - \theta)$
(note that, for the sake of the discussion, in the above equation we explicitly show the 
dependence of $\mathcal{H}$ on $\theta$, as well). Thus, we infer that, if $\mathcal{H}_{BdG}(\varphi ; \theta )$
has an energy eigenvalue $E_n$, then $\mathcal{H}_{BdG}(-\varphi , - \theta)$ must have an energy 
eigenvalue with the same energy (but opposite values of the parameters $\varphi$ and $\theta$). 
As a result, at zero temperature, the groundstate energy of the system must be invariant under 
$ ( \varphi , \theta ) \to ( - \varphi , - \theta )$ and, accordingly, once taking the derivative 
of the groundstate energy with respect to $\varphi$, one obtains $ I ( \varphi , \theta ) = - 
I ( - \varphi , - \theta )$. Setting $\theta = 0$ (which is equivalent to assuming that 
the magnetization is perpendicular to the SO field), we eventually obtain that $\varphi_0 = 0$ 
for $\theta = 0$.)  Rotating the magnetization towards the SO field results in two effects:\\
{\em i}) the appearance of  an anomaly in the CPR ($\varphi_0 \neq 0$), \\
{\em ii}) the appearance of a nonzero asymmetry ($I_{c+}- I_{c-} \neq 0$).

\begin{figure}[ht]
\includegraphics[width=1\linewidth]{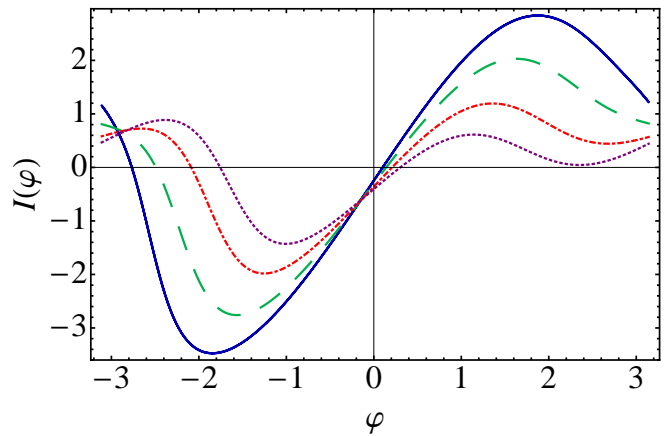}
\caption{Plots of   $I(\varphi)$ in units of $e\Delta_0/\hbar$ as a function of $\varphi$. 
Energies are measured in units of the harmonic confinement energy $E_{\omega}=\hbar \omega$ and $E_F=1.7 E_\omega$. We set the SOI to
be $\alpha=0.9 \hbar^2/m l_\omega$, and $L_{SO}$ and $L_{F}$, respectively the length of the region with SOI and the ferromagnetic 
region, equal to $l_\omega=\sqrt{\hbar/m \omega}$.The dimensionless magnetization $h'=h/E_\omega$  is 0.6 (solid curve), 
0.75 (dashed curve), 0.9 (dash-dotted curve), 1.0 (dotted curve). The angle $\theta$ is set to $\pi/2$.}
\label{current-phase-1}
\end{figure}

\begin{figure}[ht]
\includegraphics[width=0.9\linewidth]{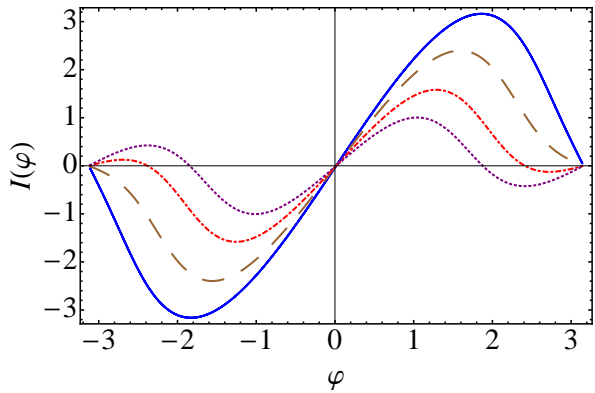}
\caption{Plots  $I(\varphi)$ in units of $e\Delta_0/\hbar$ as a function of  $\varphi$, for 
$h'=$0.6 (solid), 0.75 (dashed), 0.9 (dash-dotted), 1.0 (dotted). All the other parameters are the same as in Fig.~\ref{current-phase-1}, 
but now  $\theta=0$. Notice that in this case  $\varphi_0$ is always equal to zero.}
\label{current-phase-2}
\end{figure}

In general, discontinuities may appear in the plots of $ I ( \varphi )$ vs. $\varphi$, wich are typically 
due  to   crossings between  Andreev levels. Nevertheless, for the sake of the presentation,   in 
Figs.~\ref{current-phase-1},\ref{current-phase-2} we have chosen a set of parameters such that no discontinuities 
appear in the  CPR. Also, we stress that, for both cases in Fig.~\ref{current-phase-1} and Fig.~\ref{current-phase-2}, higher values 
of the exchange field  correspond to a smaller amplitude of the Josephson current accompanied by faster oscillations as 
a function of   $\varphi$. To explain those features, we note that, on the one hand, the reduction in the amplitude 
can be ascribed to the effect of the magnetic region which acts as spin filter, effectively reducing the 
transmission of one spin species and consequently reducing the efficiency of Cooper  pair  transfer
between the two superconducting leads. On the other hand, 
the appearance of high order harmonics in the CPR for higher values of the exchange field  appears to be 
a  precursor of a $0-\pi$ transition.

\begin{figure}[ht]
\includegraphics[width=.9\linewidth]{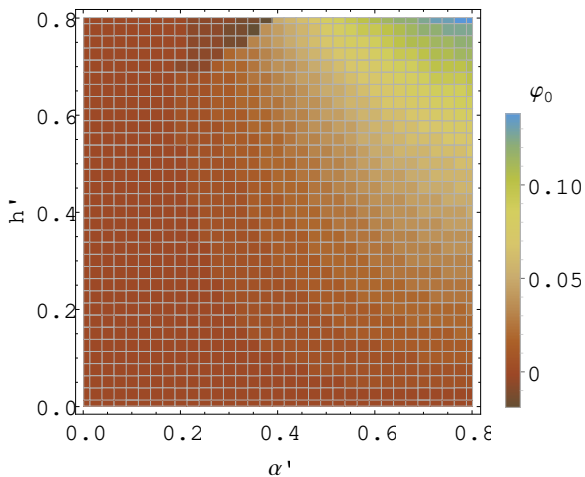}
\caption{Density plot of   $\varphi_0$  as a function of the dimensionless SOI $\alpha'=\alpha m l_\omega/\hbar^2$ and 
the dimensionless magnetization $h'=h/E_\omega$ in the regions $N_{SO}$ and $N_{F}$ respectively. 
The remaining parameters are the same as in Fig.~\ref{current-phase-1}.}
\label{2-reg-phi0}
\end{figure}

\begin{figure}[ht]
\includegraphics[width=.9\linewidth]{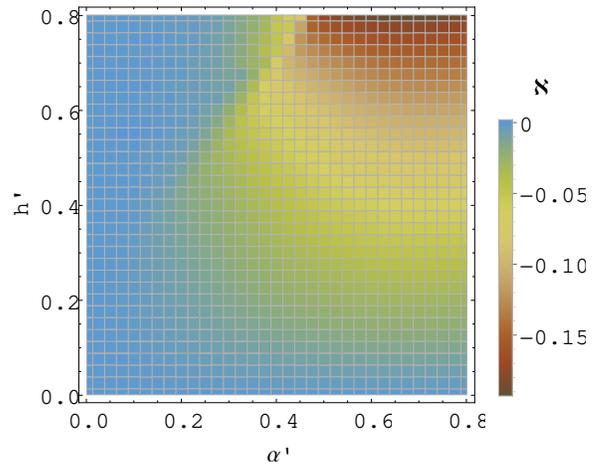}
\caption{Asymmetry of the critical current in the two direction calculated 
as a function of the dimensionless SOI $\alpha'=\alpha m l_\omega/\hbar^2$ and 
the dimensionless magnetization $h'=h/E_\omega$ in the regions $N_{SO}$ and $N_{F}$ respectively. 
The remaining parameters are the same as in Fig.~\ref{current-phase-1}.}
\label{2-reg-Ip-Im}
\end{figure}

To evidence how  $\varphi_0$ depends on the system parameters, 
in Fig.~\ref{2-reg-phi0} we show  $\varphi_0$, defined as the phase at which the  Josephson energy is minimum (and, 
accordingly, $ I ( \varphi ) = 0$), calculated    for several values of the SOI of $N_{SO}$ and of the 
exchange field of $N_{F}$.  The parameters employed to generate the plots are reported in the figure's caption.
To perform a similar analysis for the asymmetry, we therefore use the quantity 
$\aleph=(I_{c+}-I_{c-})/(I_{c+}+I_{c-})$ and, plot   $\aleph$ as a function the SOI and 
the exchange field  in Fig.~\ref{2-reg-Ip-Im}. As a main comment, it is worth pointing out that 
 $\aleph = 0 $  when the angle  between the exchange field  and the SO field is zero.
 As it is evident from Fig.~\ref{2-reg-phi0} and Fig.~\ref{2-reg-Ip-Im}  larger values of the SOI and the exchange field correspond to larger values of $\varphi_0$ and $\aleph$, if the exchange field  
is properly oriented with respect to the spin-orbit field. In order to study whether it is possible to enhance 
$\varphi_0$ and $\aleph$ without resorting to larger values of the fields in the next section we analyze
a multi-layer setup with two spin-orbit coupled and two ferromagnetic regions.

\subsection{4 Regions}
\label{results_B}

\begin{figure}[ht]
\includegraphics[width=.8\linewidth]{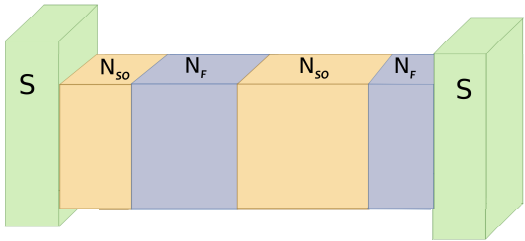}
\caption{Schematic representation of the device, in this case the normal region consists of sandwich-like structure with spin-orbit 
coupled sections ($N_{SO}$) connected to a ferromagnetic ones ($N_F$). 
The magnetization (or alternatively a magnetic field) is assumed to be in the $xy$ plane.}
\label{scheme_4_reg}
\end{figure}

We now move to discuss the setup represented in Fig.~\ref{scheme_4_reg}, in which  
the normal region consists of four different sections, with alternating  SOI coupled and 
ferromagnetic regions.
Again we calculate the scattering matrix of each region and then after translating them to the proper position 
we construct the full scattering matrix,  as previously explained.  In order to compare the results of this 
subsection to those of the previous one we, assume that the total length of the two spin-orbit coupled 
(ferromagnetic) region is equal to that of the single spin-orbit (ferromagnetic) region in the two-region setup. 
In this way we can assess whether, and to what extent, increasing the number of layers works to 
maximize the anomaly in the CPR, as well as to recover a larger values of
$\aleph$, that is,  to obtain a larger superconducting rectifying affect.

Here, to avoid further complications, we take the orientation of the exchange field in the two ferromagnetic regions
to be along the y direction, i.e. orthogonal to the spin-orbit field. In principle one might allow for different orientation 
of the exchange field in the two ferromagnetic regions but, possibly, the case addressed below corresponds to the
most accessible configuration in real devices. 
Notwithstanding the difficulty of orienting the exchange field in the two ferromagnetic regions, in light of the results of
Ref.~\onlinecite{Silaev:2017} it would be reasonable to expect that larger values of $\varphi_0$ and $\aleph$ can be obtained by fine tuning 
the angle between the magnetizations.

\begin{figure}[ht]
\includegraphics[width=.9\linewidth]{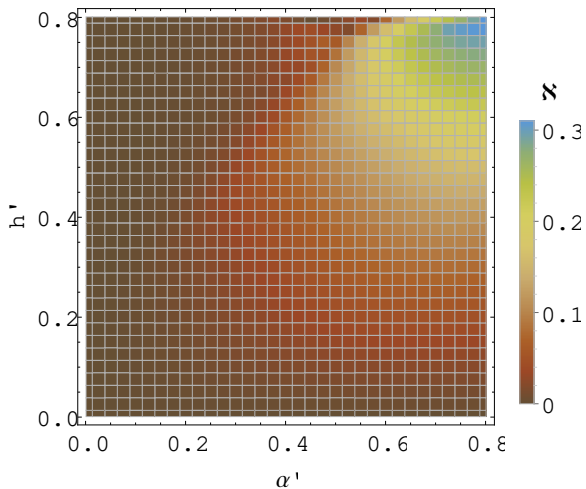}
\caption{Asymmetry of the critical current in the two direction for the four regions set-up,  calculated 
as a function of the dimensionless SOI $\alpha'=\alpha m l_\omega/\hbar^2$ and 
the dimensionless magnetization $h'=h/E_\omega$ in the regions $N_{SO}$ and $N_{F}$ respectively.  For the numerical calculation we have set $L_{SO1}=L_{F1}=0.35~l_{\omega}$, 
and $L_{SO2}=L_{F2}=0.65~l_{\omega}$. The remaining parameters are the same as in Fig.~\ref{current-phase-1}.}
\label{4-reg-Ip-Im}
\end{figure}

By analyzing the CPR for several values of the spin-orbit coupling and the magnetization, 
as well as changing the relative magnitude of $L_{so1}$ ($L_{F1}$) and $L_{so2}$ ($L_{F2}$), we find
the that the magnitude of $\varphi_0$ is in general of the same order of magnitude for the two- and the four-regions setups. 
Conversely, and most importantly for future applications, we find that the asymmetry $\aleph$ between $I_{c+}$ 
and $I_{c-}$  in the four-regions setup can be enhanced to the two-regions one by an asymmetric choice 
of the lengths of the different sections. Indeed, as we show in Fig.~\ref{4-reg-Ip-Im} in the four region setup for the 
regions of parameters considered, we find maximum values of $\aleph\simeq0.3$, whereas for the 
two region setup we obtain at most $\aleph\simeq0.15$. This result would suggest that multilayer 
heterostructure as the one studied here may be useful in designing rectifying superconducting devices.

\section{Random matrix analysis}
\label{ramaan}

In this section, we discuss how our results about AJE and nonzero asymmetry are related to   the total number of open channels 
that we take into account. As we pointed out before,  we are interested in setting the system parameters so to maximize both
$\varphi_0$ and the critical current asymmetry. By direct calculation 
(not illustrated here),   we found $I_{c+}=I_{c-}$ when  $N=1$, while, to find   $ I_{c+} - I_{c-} \neq 0 $, 
we have to set   $N\ge2$. To the best of our knowledge, there is no {\em a priori } 
reason why only for $N=1$ one should have $I_{c+}=I_{c-}$. Thus,  in order to understand whether 
this finding is  accidental to our model, or it rather  occurs in general we have performed a 
numerical simulation using random scattering matrices  to describe the normal region (note that 
resorting to random scattering matrices is a standard mean to deal, for instance, with 
dephasing effects in mesoscopic systems \cite{been_random,lgt}). Specifically, 
we assume that the two superconducting leads are connected to each other 
by a normal region characterized by a scattering matrix 
$\hat{S}$. We take $\hat{S}$ to be a unitary matrix, whose elements are extracted 
with an uniform probability distribution,  with no further restriction. 
Since we look for Josephson junctions which exhibit  anomalies in the Josephson CPR, we do not enforce  
symmetries on the matrix $S$, such as  time reversal, or spin-rotational 
symmetry\cite{Beenakker:1997}: in fact,  in the presence of either one of these latter symmetries (or of both
of them), $\hat{S}$ 
would belong respectively to the orthogonal and to the symplectic group. Using symmetry arguments
it can be shown that for these two symmetry classes $\varphi_0=0$ \cite{Yokoyama:2013,Campagnano:2015}. 
By means of Eq.~\ref{determinant},  we therefore calculate the Andreev spectrum and the Josephson current, 
computing then  $I_{c+}$ and $I_{c-}$ for each random realization.
To quantify the asymmetry, we  use the mean square visibility $\langle \aleph^2 \rangle$, with 
$\aleph = ( I_{c+} - I_{c-} ) / ( I_{c+} + I_{c-} )$ for a given scattering matrix, and $\langle \ldots \rangle$ denoting 
the average over a large number of different realization of the random matrix.  We repeat the calculation for the number of 
open transport channels $N=1,2,3,4$. For each case we generate
$\mathcal{N}=50000$ random scattering matrices and, using Eq.~\ref{determinant}, we compute the Andreev spectrum and the Josephson current and,
eventually, we compute  $\aleph=(I_{c+}-I_{c-})/(I_{c+}+I_{c-})$ for each realization of $\hat{S}$. 

\begin{figure}[ht]
\includegraphics[width=.95\linewidth]{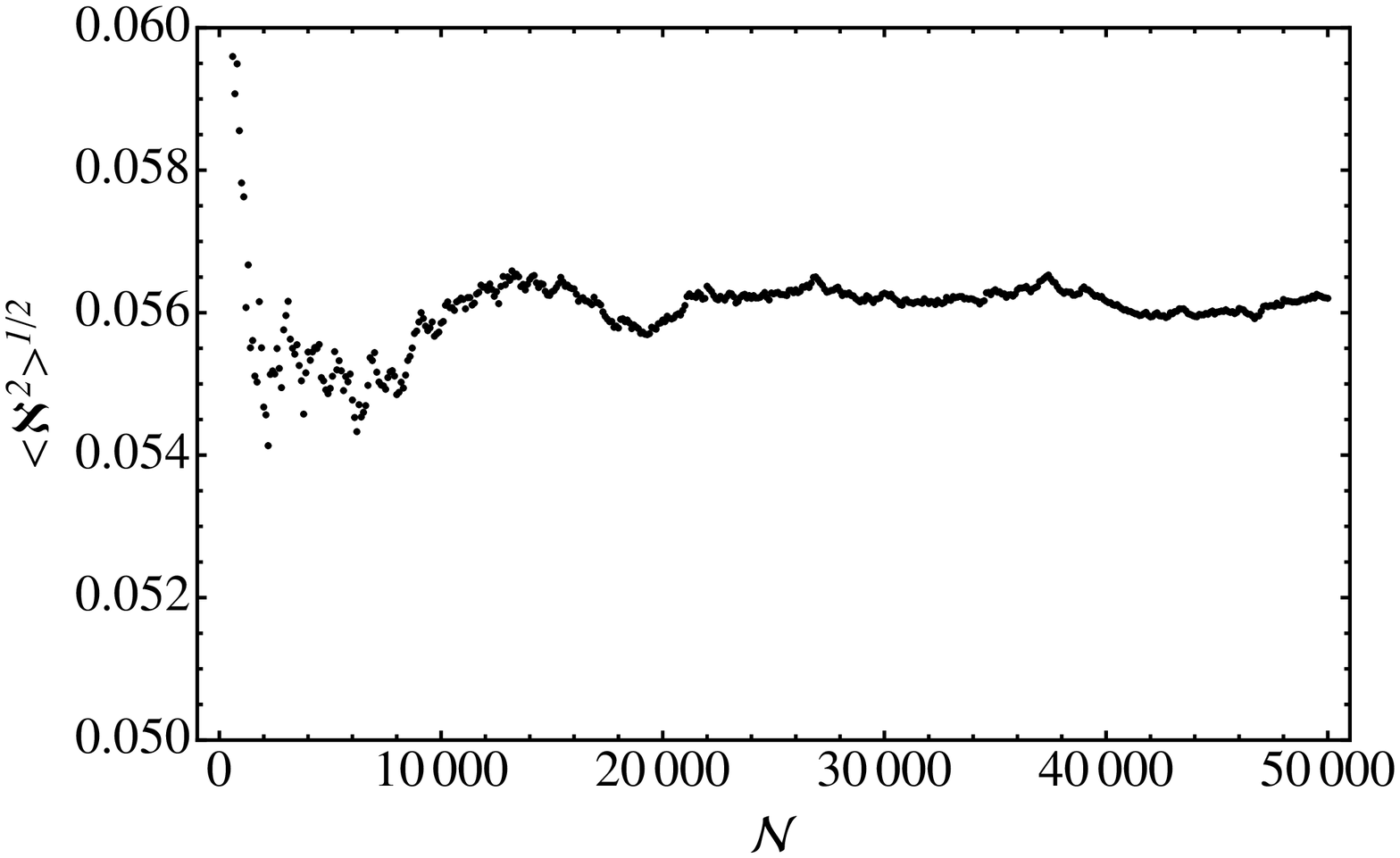}
\includegraphics[width=.95\linewidth]{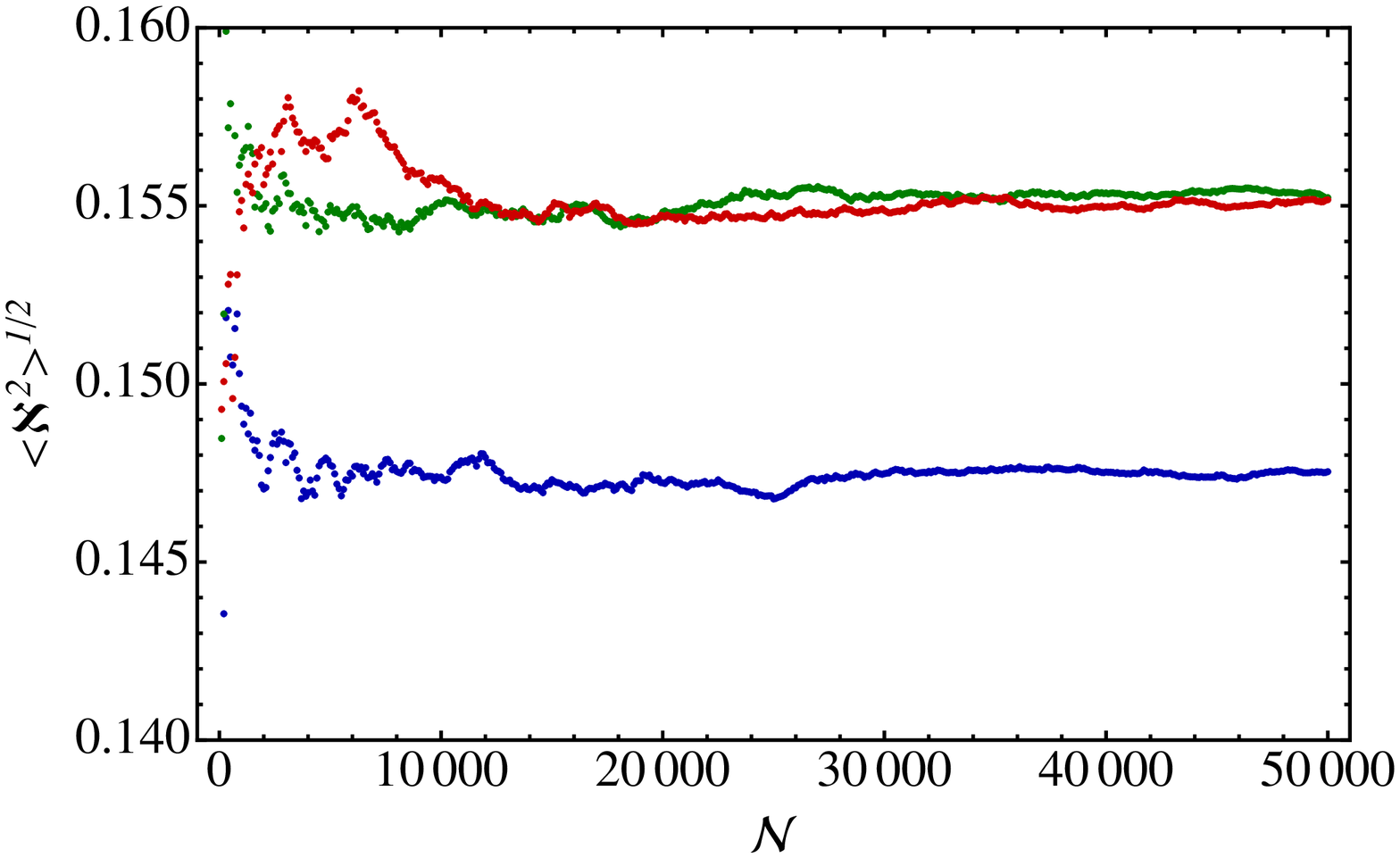}
\caption{Random matrix analysis: We generate random unitary scattering matrices $\hat{S}$ and calculate the 
critical current asymmetry $\aleph$, for $N=1,2,3,4$ 
open transport channels, each having two spin directions. We generate $\mathcal{N}=50000$ 
matrices for each case and and plot $\sqrt{\langle \aleph^2 \rangle}$ vs the 
number of realizations. Upper panel: $N=1$. Lower panel: $N=2$ blue curve, $N=3$ green curve, $N=4$ red curve.}
\label{rmt}
\end{figure}
In Fig.~\ref{rmt} we plot the computed value of $\sqrt{\langle \aleph^2 \rangle}$ as a function of the number of realizations. 
We find that $\sqrt{\langle \aleph^2 \rangle}\sim0.05$ for $N=1$, so that only a small asymmetry can be observed in this case, 
in accordance with our calculation using the Hamiltonian of Eq.~\ref{hamiltonian}. Moreover, non-zero values of $\aleph$ are 
found only for realization of $\hat{S}$ such that the CPR is discontinuous. For the case 
$N\ge2$ we find $\sqrt{\langle \aleph^2 \rangle}\sim0.15$ and the asymmetry can be observed even for a continuous CPR.
It should be stressed that, within the approach presented here, we are properly describing the 
Josephson effect through a cavity and not the case of a wire\cite{Beenakker:1997,Campagnano:2006}; 
the latter case will be the subject of a further study.

\section{Summary and Outlook}
\label{summary}

In this work we have demonstrated that the anomalous Josephson effect can be expected in SNS junctions where the normal region in a 
heterostructure formed by alternating ferromagnetic and spin orbit coupled segments. 
We have shown that when the Fermi energy is such that the number of transport channels $N\ge 2$ it is possible to observe a sizeable direction 
dependency of the critical current; we have validated this result also using a random matrices analysis. Moreover we have shown that the asymmetry 
between $I_{c+}$ and $I_{c-}$ can be enhanced using a four layer heterostructure vs a two layer one. Our findings might be relevant to the design of 
devices with large  $I_{c+}$, $I_{c-}$ asymmetry to be employed as diodes in superconducting circuits.

\appendix

\section{Calculation of the Scattering matrix}
\label{scatmay}

In this appendix, we outline the calculation of the $\hat{S}_{e}(0)$-matrix for the normal region. In view of the 
relation $ \hat{S}^*_{h}(0)=\hat{S}_{e}(0)$, by means of the same token, we 
compute the $\hat{S}_{h}(0)$-matrix for the normal region, as well. 
In practice, we first  divide the normal region in a SOI and a ferromagnetic segment
and separately derive the scattering matrices of the two regions, respectively referred 
to in the following as   $S_{SO}$ and $S_{F}$. Eventually, we combine the two 
of them to calculate the  $\hat{S}_{e}(0)$-matrix for the whole normal region.
In fact, apart for the technical subtelties in combining together  $\hat{S}_{SO}$ and $\hat{S}_{F}$, 
our approach appears to be particulary convenient, as  it  
  allows us to  generalize our study to multi-layer setups. To combine together $\hat{S}_{SO}$ and $\hat{S}_{F}$, 
  it is more convenient to resort to the transfer 
matrices, for which a simple composition rule exists. To do so, we decompose each 
scattering matrix $\hat{S}_{\rho}$ $(\rho=SO,F)$ into  reflection and transmission blocks,
according to 

\begin{equation}
\hat{S}_\rho = 
\left(\begin{array}{cc}
\hat{r}_\rho &  \hat{t}_\rho'\\
\hat{t}_\rho    & \hat{r}_\rho'  
\end{array}
\right)
\: . 
\label{scatblocks}
\end{equation}
\noindent
Next, we  introduce the transfer matrices $\hat{M}_{SO}$ and $\hat{M}_{F}$. 
By definition, each $\hat{M}_\rho$ relates the scattering amplitudes to the left-hand side of 
the corresponding region to the ones at the right-hand side, according to 
 
\begin{equation}
\left(\begin{array}{c}
b_{\rho,eR} \\
a_{\rho,eR}    
\end{array}
\right)
=\hat{M}_\rho
\left(\begin{array}{c}
a_{\rho,eL} \\
b_{\rho,eL}    
\end{array}
\right) \,,
\end{equation}
with $\left\{a_{\rho,eL},a_{\rho,eR},b_{\rho,eL},b_{\rho,eR}\right\}$ denoting the 
scattering amplitudes across the corresponding   scattering regions.
In analogy with the scattering matrices, the transfer matrices  admit 
a block decomposition, as well, according to 

\begin{equation}
\hat{M}_\rho = 
\left(\begin{array}{cc}
\hat{m}_{\rho,11} &  \hat{m}_{\rho,12} \\
\hat{m}_{\rho,21}     & \hat{m}_{\rho,22}  
\end{array}
\right) \,.
\label{transblock}
\end{equation}
The blocks in Eq.~\ref{scatblocks} and in Eq.~\ref{transblock} are related to each other 
according to the relations
 
\begin{eqnarray}
\hat{m}_{\rho,11}=\hat{t}_\rho^{\dag-1}, &   \hat{m}_{\rho,12}=\hat{r}'_{\rho}\hat{t}_{\rho}'^{-1}   \nonumber  \\
\hat{m}_{\rho,21}=-\hat{t}_{\rho}'^{-1}\hat{r}_{\rho}, & \hat{m}_{\rho,22}=\hat{t}'^{-1}_\rho \, ,
\label{transfer-matrix-blocks}
\end{eqnarray}
\noindent
together with their inverse 

 \begin{eqnarray}
\hat{t}_\rho= \hat{m}_{\rho,11}^{\dag-1}  , &   \hat{r}_\rho=  -\hat{m}_{\rho,22}^{-1}\hat{m}_{\rho,21} \nonumber  \\
\hat{t}'_\rho= \hat{m}_{\rho,22}^{-1}  , &  \hat{r}'_\rho = \hat{m}_{\rho,12} \hat{m}_{\rho,22}^{-1}    \, .
\label{scattering-matrix-blocks}
\end{eqnarray}
\noindent
To derive the transfer matrix, we separately solve the Schr\"odinger equation in the 
various normal regions by setting $x=0$ at the center of each region. 
Eventually, using the composition law of the transfer matrices, we shift the 
corresponding matrices according to their
location within the heterostructure and combine them to obtain the total transfer matrix as  $\hat{M}=\hat{M}_{F}\hat{M}_{SO}$. 
From the total transfer matrix we then calculate the full scattering matrix, which we use to compute the Andreev spectrum.

\subsubsection{Scattering matrix of spin-orbit coupled region} 
\label{scamaspi}

We begin our calculation by deriving  $\hat{S}_{SO}$. To do so, by standard methods, we explicitly solve the 
Schr\"odinger equation in the spin-orbit region and at its left- and right-hand side,  where only transverse confinement is assumed.
Eventually, we  match the solutions at the interfaces. When doing the corresponding calculations, 
we let the Fermi energy vary in an interval such that only two transport channels are open, 
each with two spin orientations. 

The wave functions corresponding to the 
scattering states at energy $E$ to the left- and to the right-hand side  of the 
SOI region can be readily written as

\begin{equation}
\psi_{n\sigma,L/R}(E;x,y)=e^{\pm i k_n x}\chi_n(y)\phi_\sigma \, ,
\label{scatsoi}
\end{equation}
\noindent
with $n=1,2$ and the $\pm$-sign referring to the  right-going and to the left-going states.
In Eq.~\ref{scatsoi},  $\chi_n(y)$ and $\phi_\sigma$ are respectively the eigenfunctions of 
the harmonic oscillator and of the spin Pauli matrix $\sigma_z$. In particular, we label  the 
groundstate of the harmonic oscillator with   $n=1$, the first excited state with $n=2$, and so on. 
Moreover, we set $k_1=[2m (E-\hbar \omega/2)]^{1/2}/\hbar$ and 
$k_2=[2m (E-3\hbar \omega/2)]^{1/2}/\hbar$. At variance, for $n>2$, 
there are no propagative solutions and the corresponding (evanescent) modes 
are described by the wave functions

\begin{equation}
\psi_{n\sigma,L/R}(E;x,y)=e^{\pm  \kappa_n x}\chi_n(y)\phi_\sigma \, ,
\end{equation}
with $\kappa_n=\{2m [(2n-1)\hbar \omega/2-E])\}^{1/2}/\hbar$, where the + (-) sign refers to the left-hand 
side (right-hand side) region.

To obtain the eigenfunctions in the SOI region, we numerically diagonalize the Hamiltonian using the basis $\{e^{i \kappa x} \chi_m(y)\phi_\sigma\}$.
For simplicity we truncate the Hilbert space considering the first three sub-bands, resulting in a $6\times6$ Hamiltonian matrix. 
Such an approximation is expected to   give a reasonable description of the system even in the presence of a sizeable SOI\cite{Governale:2002}.
Accordingly, our problem is now reduced to finding the eigenvalues 
and the eigenfunction of the corresponding $6\times6$-finite dimensional
Hamiltonian matrix
$\mathcal{H}(\kappa)_{m,\sigma;m',\sigma'}$. For a given energy E, 
the allowed $\kappa_i$ are obtained from the equation 
\begin{equation}
{\rm det} \left[ \mathcal{H}(\kappa_i)-E \right]=0 \; , 
\end{equation}
\noindent 
which implies that, for each value of the energy we have 12 solutions  $\kappa_i$ $(i=1,..,12)$,
with the corresponding eigenfunctions given by:

\begin{equation}
\psi_{so}(x,y;E)=\sum_{i=1}^{12} b^{so}_i e^{i \kappa_i x}\sum_{m \sigma} c^{(i)}_{m, \sigma} \chi_{m}(y)\phi_\sigma
\label{psiSO}
\end{equation}
where the coefficients $c^{(i)}_{m,\sigma}$ have to be determined numerically, while the coefficients $b^{so}_i$ are
determined by imposing the proper matching conditions, as discussed  below. For  two open transport channels in the 
leads, each one with both spin polarizations,  the electronic scattering matrix $\hat{S}_{SO}$ takes the form:  
\begin{equation}\label{S-blocks}
\hat{S}_{SO}=\left (
\begin{array}{cc}
\hat{r} & \hat{t}' \\
\hat{t} & \hat{r}'
\end{array}
\right),
\end{equation}
with the ($4 \times 4$) block $\hat{r}$ given by 

\begin{equation}
\hat{r}=\left (
\begin{array}{cccc}
r_{1\uparrow,1\uparrow} & r_{1\uparrow,1\downarrow} &  r_{1\uparrow,2\uparrow}  &r_{1\uparrow,2\downarrow}\\
r_{1\downarrow,1\uparrow} &r_{1\downarrow,1\downarrow} &r_{1\downarrow,2\uparrow} &r_{1\downarrow,2\downarrow}\\
r_{2\uparrow,1\uparrow}  &r_{2\uparrow,1\downarrow} & r_{2\uparrow,2\uparrow} & r_{2\uparrow,2\downarrow} \\
r_{2\downarrow,1\uparrow} &r_{2\downarrow,1\downarrow}&r_{2\downarrow,2\uparrow} &r_{2\downarrow,2\downarrow}
\end{array}
\right),
\end{equation}
and similar expressions for $\hat{r}', \hat{t}$ and $ \hat{t}'$. 
To move ahead in the calculation,   one has to compute  all the reflection and transmission 
coefficients, by matching the wave function in Eq.~\ref{psiSO} with the one in the leads, 
for any possible choice of scattering boundary conditions. To illustrate how the procedure works,  
let us explicitly discuss  the case of a spin-up  particle incoming from the left-hand side. 
In this case, the wave functions within the left-hand side ($L$) and the right-hand side ($R$)
are respectively given by:
\begin{multline}
\psi_L(x,y;E)=e^{i k_1 x} \chi_{1}(y)\phi_\uparrow+r_{1 \uparrow,1 \uparrow}e^{-i k_1 x}\chi_{1}(y)\phi_\uparrow \\
+ r_{1 \downarrow,1 \uparrow}e^{-i k_1 x}\chi_{1}(y)\phi_\downarrow   
+r_{2 \uparrow,1 \uparrow}e^{-i k_2 x}\chi_{2}(y)\phi_\uparrow 
\\+ r_{2 \downarrow,1 \uparrow}e^{-i k_2 x}\chi_{2}(y)\phi_\downarrow+  
 \sum_{\sigma=\uparrow, 
\downarrow} d^{L}_{\sigma} \,e^{\kappa_3 x}\chi_{3}(y)\phi_\sigma \, ,
\end{multline} 
\begin{multline}
\psi_R(x,y;E)=t_{1 \uparrow,1 \uparrow}e^{i k_1 x} \chi_{1}(y)
\phi_\uparrow+t_{1 \downarrow,1 \uparrow}e^{i k_1 x}\chi_{1}(y)\phi_\downarrow \\
t_{2 \uparrow,1 \uparrow}e^{i k_2 x} \chi_{2}(y)
\phi_\uparrow+t_{2 \downarrow,1 \uparrow}e^{i k_2 x}\chi_{2}(y)\phi_\downarrow \\
+\sum_{\sigma=\uparrow,\downarrow} d^{R}_{i,\sigma} \,e^{-k_3 x}\chi_{3}(y)\phi_\sigma \,.
\end{multline} 
Let us denote with  $L_{SO}$ the total length of the SOI region and, to simplify the 
derivation, let us assume that the interfaces are symmetrically located at  $x=\pm L_{SO}/2$.
The matching conditions at the interfaces require that the wave function is continuos while, in 
general,  its derivative with respect to $x$  must be discontinuous, to account  for the discontinuous SOI 
interaction (cfr. Eq.~\ref{hamiltonian}). 
Projecting the equations corresponding to the  matching conditions onto the basis states $\chi_m(y)\phi_\sigma$
($m=1,...,l;\sigma=\uparrow,\downarrow$) we obtain the following set of equations:

\begin{equation}
\int_{-\infty}^{+\infty}\chi_m^*(y)\phi_\sigma^{\dag}\left[\psi_L(- L_{SO}/2,y)-\psi_{so}(- L_{SO}/2,y)\right]dy=0,
\end{equation}
\begin{equation}
\int_{-\infty}^{+\infty}\chi_m^*(y)\phi_\sigma^{\dag}\left[\psi_R( L_{SO}/2,y)-\psi_{so}( L_{SO}/2,y)\right]dy=0.
\end{equation}
\begin{multline}
\int_{-\infty}^{+\infty}\chi_m^*(y)\phi_\sigma^{\dag}\Big\{\partial_x \psi_{so}(- L_{SO}/2,y)-\partial_x 
\psi_{L}(- L_{SO}/2,y) \\ -
\frac{i m}{\hbar^2}\alpha \sigma_y \psi_{so}(- L_{SO}/2,y) \Big\}dy=0,
\end{multline}
\begin{multline}
\int_{-\infty}^{+\infty}\chi_m^*(y)\phi_\sigma^{\dag}\Big\{\partial_x \psi_R( L_{SO}/2,y)-\partial_x \psi_{so}( L_{SO}/2,y) \\ +
\frac{i m}{\hbar^2} \alpha \sigma_y \psi_{so}( L_{SO}/2,y) \Big\}dy=0.
\end{multline}
Therefore, we have a set of $8l$ equations which we solve numerically to determine the corresponding $\hat{S}$ matrix
 elements. Repeating  the calculation for each possible incoming channel we construct   $\hat{S}_{SO}$ as a function 
of the energy $E$. Eventually, consistently with the above discussion,  we set  $E=E_F$.

\subsubsection{Scattering matrix of the ferromagnetic region}
\label{scafes}

The calculation of $\hat{S}_F$ is quite simpler, since, in this case, it is straightforward 
to explicitly solve the   Schr\"odinger equation and to find the corresponding eigenvalues and 
eigenfunctions. In the case of  in-plane magnetization, corresponding to the 
unit vector $\hat{n}=(\cos(\theta),\sin(\theta),0)$, the eigenfunctions are given by

\begin{equation}
\psi_F(x,y;E)=e^{i k_n x}\chi_n(y)\phi_{\pm} \, ,
\label{scafes.1}
\end{equation} 
with $\phi_{\pm}=(\pm \exp(-i \theta ),1)/\sqrt{2}$  spinors in the spin space,
 $k_n=\pm\sqrt{2m[E-\hbar \omega(n-1/2)\mp h_0]}/\hbar$. From the 
 wavefunctions in Eq.~\ref{scafes.1} it is now straightforward to compute $\hat{S}_F$ by 
 exactly the same procedure we have used to derive $\hat{S}_{SO}$, which is even more
 simplified by the fact that  the wave functions and their derivatives are both 
continuous at the interfaces.

\subsubsection{Translation of the Scattering potential}
\label{transpo}

For convenience,  in computing $\hat{S}_{SO}$ and $\hat{S}_F$, we have assumed that 
the corresponding regions were symmetric with respect to the  origin of the $x$-axis.  
Now, when composing the results to construct the full $\hat{S}$-matrix,   we need to translate 
the center of  scattering regions to its proper position, so that i.e.  the SO region ranges between $x_L$ and $x_c$ and
the F region between $x_c$ and $x_R$ (a pertinent generalization of such a procedure will lead us 
to correctly approach, in the following,    a sandwich-like structure with more than two regions).

To illustrate our procedure, let us consider  a scattering matrix $\hat{S}$, determined by some   potential $V$, defined so 
that  
\begin{equation}
\left(
\begin{array}{c}
b_L \\
b_R
\end{array}
\right)
=\hat{S}
\left(
\begin{array}{c}
a_L \\
a_R
\end{array}
\right) \,
\end{equation}
with the block decomposition of Eq.~\ref{S-blocks} for $\hat{S}$.
Assuming, as we have done  throughout our paper, that an equal number $N$ of transport channels 
is available at the left-hand side  and at the right-hand side of the scattering region, the blocks 
$\hat{r},\hat{r}',\hat{t},\hat{t}'$ will 
be realized as $N\times N$ matrices. Let  $\tilde{V}$ be the scattering potential obtained by translating $V$ by a distance 
d along the x-axis and let $\tilde{\psi}$ and $\psi$  be the solutions of the Schr\"odinger equation 
respectively corresponding to $\tilde{V}$ and to $V$ respectively, so that one has   $\tilde{\psi}(x+d)=\psi(x)$. 
Making use of this last relation, it is straightforward to show that the scattering matrix $\tilde{S}$ relative to $\tilde{V}$ 
can be obtained from $\hat{S}$ by the following transformation:
\begin{equation}
\tilde{S}=
\left(
\begin{array}{cc}
\Lambda(d) & 0 \\
0 & \Lambda^{-1} (d)
\end{array}
\right)
\hat{S} 
\left(
\begin{array}{cc}
\Lambda(d) & 0 \\
0 & \Lambda^{-1} (d)
\end{array}
\right) \, ,
\end{equation}
with $\Lambda(d)=\mbox{diag}(\exp(i k_1 d),...,\exp(i k_N d))$.
In terms of the blocks of the scattering matrix we have:
\begin{equation}
\tilde{S}=
\left (
\begin{array}{cc}
\Lambda(d) \hat{r} \Lambda(d) & \Lambda(d) \hat{t}' \Lambda^{-1}(d)\\
\Lambda^{-1}(d) \hat{t} \Lambda(d) &\Lambda^{-1}(d)  \hat{r}' \Lambda^{-1}(d)
\end{array}
\right) \,,
\end{equation}
and similarly for the transfer matrix  
\begin{equation}
\tilde{M}=
\left (
\begin{array}{cc}
\Lambda(d)^{-1} \hat{m}_{11} \Lambda(d) & \Lambda^{-1}(d) \hat{m}_{12} \Lambda^{-1}(d)\\
\Lambda(d) \hat{m}_{21} \Lambda(d) &\Lambda(d)  \hat{m}_{22} \Lambda^{-1}(d)
\end{array}
\right) \,.
\end{equation}


\end{document}